# Experimental Demonstration of a New Oscillator Concept Based on Degenerate Band Edge in Microstrip Circuit

Dmitry Oshmarin, *Student, IEEE*, Ahmed F. Abdelshafy, *Student, IEEE*, Alireza Nikzamir, *Student, IEEE*, Michael M. Green, *Senior Member, IEEE*, and Filippo Capolino, *Fellow, IEEE*

*Abstract*— The first experimental demonstration of an oscillator based on a periodic, resonant microstrip circuit with a degenerate band edge (DBE) is presented. The DBE is a fourth-order exceptional degeneracy of the eigenmodes in a lossless periodic waveguide that is visible in the wavenumber-frequency dispersion diagram, and the periodic microstrip behaves as a frequency selective medium. The presence of the DBE condition and the associated DBE resonance allow for a stable, single-mode oscillation as well as stability with varying the load resistance.

## I. Introduction

AN oscillator is a circuit that produces a continuous, single-tone frequency response. It is one of the fundamentally integral components of any radio frequency (RF) system, from transceivers to microscopy [1], [2]. The most common type of oscillator is the harmonic oscillator consisting of an active device, usually in the positive feedback configuration, which provides the energy to the passive portion of the oscillator that acts as the frequency selective component. The active device is required to compensate for the losses and meet the Barkhausen criteria, which is a requirement for a steady-state oscillation (but not sufficient on its own). The active device can be realized in a variety of ways such as Gunn diode [3], cross-coupled transistor pair [4]–[8], op-amp [9], and a multitude of other ways [10], [11]. The focus of this paper is to demonstrate a first-ever practical realization of a degenerate band edge (DBE) oscillator. In particular, this paper presents an oscillator based on a spatially periodic structure with a $4^{th}$ order degeneracy of the eigenstates, consisting of cascaded unit cells, that act as the frequency selective medium for the oscillation.

The periodic structure used in this paper, as well as any periodic structure able to support an electromagnetic wave, can be described by the evolution of the voltages and currents, or eigenmodes, from one cell to the next. When these eigenmodes coalesce into a single eigenmode, a degeneracy of eigenmode is formed [12]–[14]. The point in frequency and wavenumber at which this degeneracy forms is known as the exceptional point of degeneracy (EPD), the order of which is described by the number of coalescing eigenmodes at the band edge. The focus of this paper is on a $4^{th}$-order EPD, with four coalescing eigenmodes, known as DBE that exists in lossless periodic structures [13], [15], [16]. Near the band edge, the dispersion relation, which relates frequency to wavenumber, is characterized as $(\omega_d - \omega) \propto (k - k_d)^4$, where $\omega_d$ is the angular frequency at which DBE condition occurs, with subscript $d$ denoting DBE, $k$ is the Floquet-Bloch wavenumber, and $k_d = \pi/d$ is the DBE wavenumber at the edge of Brillouin zone, with $d$ being the length of a unit cell [13], [15], [17]. The exponent of 4 is an indication of a fourth-order degeneracy. A finite length periodic structure, such as the one studied and realized in this paper, will form a Fabry-Perot cavity with a multitude of resonances, with any of those resonances being a possible frequency of oscillation [18]. We are interested in the resonance closest to the DBE frequency, denoted as $\omega_{d,r}$. Operating at $\omega_{d,r}$ provides properties most desired for an oscillator, such as an enhanced quality factor, stability of oscillation despite the presence of perturbations, and stability with respect to a large range of load variations. This type of DBE based oscillator has been studied before [19]–[21], but never fabricated and validated, which is the goal reported in this paper.

A periodic structure able to support a DBE resonance is very versatile in its design and can be realized in a variety of ways, such as lumped elements [22], [23], optical waveguides [12], [24]–[26], as well as transmission lines [27]–[31], including experimental demonstrations in circular metallic waveguides [32], [33] and printed microstrip lines [29], [31]. The focus of this paper is the realization of an oscillator using a DBE in microstrip technology due to its excellent transmission characteristics in the RF range, ease of use, and general versatility based on the application. A schematic representation of the periodic structure with DBE resonance along with the negative resistance active device (A.D.) is shown in Fig. 1(a). In Sec. II, we introduce the unit cell of the periodic structure and show its passive behavior in the wavenumber-frequency domain. In Sec. III, we demonstrate the measured passive resonance behavior of the periodic structure comprising *N* unit cells. In Sec. IV, the behavior of the chosen active device is shown along with its implementation in the oscillator circuit. In Sec. V, we demonstrate and characterize the new kind of oscillator based on the DBE.

## II. DBE Resonance in Passive Microstrip Structure

The oscillator concept of Fig. 1(a) is based on two parallel microstrip lines with periodic microstrip coupling as shown in Fig. 1(b) along with its dimensions. We fabricate a unit cell on a Rogers RO3003 substrate with 0.76 mm (30 mils) thickness and dielectric constant of 3. The upper line was designed to have 50 Ω characteristic impedance.

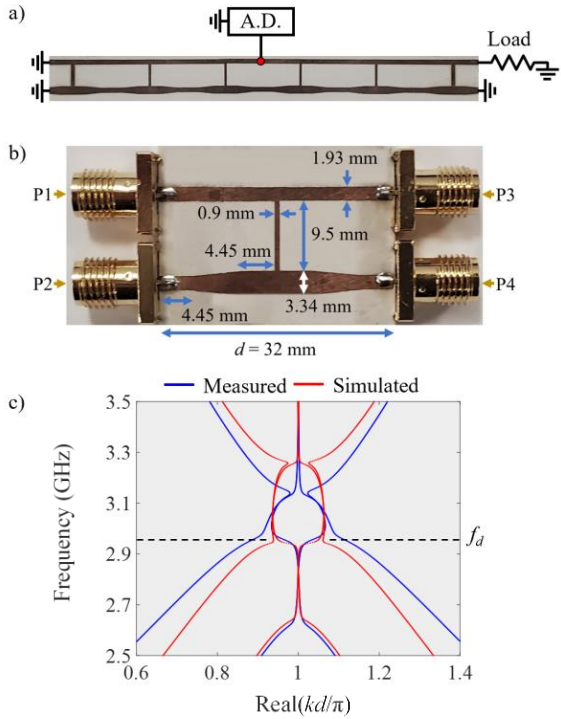

Fig. 1. (a) Schematic representation of a DBE oscillator based on periodic coupled microstrips over a grounded dielectric substrate, and a negative resistance active device (A.D.) at the center of the cavity. (b) Unit cell of the periodic microstrip circuit (the gray area is the top view of the grounded substrate) along with its dimensions. (c) Dispersion diagram of the modes in the fabricated and simulated structures showing agreement between the two results and the DBE frequency $f_d$ around 2.93 GHz, visible from the coalescence of four branches of the dispersion.

Figure 1(c) shows the dispersion diagram of the eigenmodes in the structure of Fig. 1(b), for both simulated and measured results. This provides the relation between the applied frequency and wavenumber in an infinitely long periodic structure of repeated unit cells. The simulated results were obtained from the Method of Moments implemented in Keysight ADS Method of Moments Momentum Microwave simulator. The experimental results are obtained using the fabricated unit cell using a Keysight E5071C Network Analyzer. In both cases, a 4×4 scattering (S)-parameter matrix was extracted and using MathWorks MATLAB converted to a transfer matrix. Then, the dispersion diagram is obtained by evaluating the eigenvalues derived from the transfer matrix at each frequency and converted to Bloch wavenumbers (see [22], [29] for an in-depth analysis). The real part of wavenumbers versus frequency, for both the measured and simulated results, is plotted in Fig. 1(b), which shows reasonable agreement between the two methods. While there are some differences in the dispersion diagram between the simulated and measured results, it can be seen that the four distinct modes, two propagating and two evanescent, just below the frequency $f_d$, nearly coalesce into a single, degenerate mode at the angular frequency $f_d$= 2.93 GHz, developing the aforementioned DBE condition. Above $f_d$, i.e., in the bandgap, a signal would experience strong attenuation for frequencies up to the upper bandedge (frequency point at which four modes nearly coalesce again), which is 3.15 GHz for measured and 3.3 GHz for simulated results (inside the bandgap, the modes are "evanescent"). In other words, in the bandgap frequency range, the imaginary part (not shown here) of all the four modes is different from zero indicating strong attenuation. Moreover, the imaginary parts of the two propagating modes below $f_d$ are nearly zero, and the imaginary parts of the two evanescent modes below $f_d$ are different from zero. The frequency shift between measured and simulated results can be due to various factors such as fabrication imperfections, VNA calibration issues, or numerical errors during the full-wave simulations. In general, a structure like this possesses many exceptional points of degeneracy, which are more sensitive to losses with increasing frequency. The described unit cell serves as a building block for the oscillator design discussed in the next section.

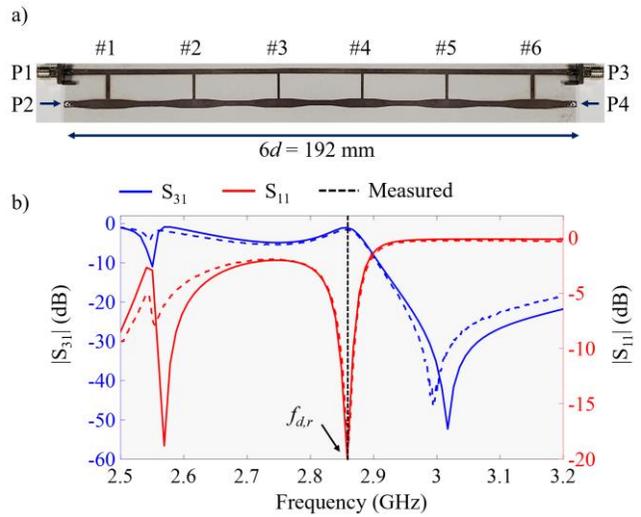

Fig. 2. (a) Periodic coupled microstrips with 6 unit cells and corresponding ports labeled P1 through P4. P2 and P4 are grounded. (b) Transmission and reflection coefficients from Port 1 to Port 3 showing the resonances at $f_{r,d}$ near the DBE frequency $f_d$ with both Ports 2 and 4 grounded.

Before the transient analysis is performed on the structure made of $N$ unit cells in tandem, passive behavior must be considered to gain a better understanding of how to incorporate active device/devices among other things. We use the analysis that closely parallels the work and studies done in [19], [20], and thus some of the finer details of the analysis will be omitted for brevity and can be found in the References. Fig. 2(a) shows the fabricated 6-cell structure. Ports P1 and P3 are used in the measurement setup, while Ports P2 and P4 are grounded. The measured 2×2 S matrix is related to Ports P1 and P3. The number of cells was carefully chosen based on the material losses of RO3003 substrate among other considerations. Since the DBE is very sensitive to losses and perturbations [29], [34], every unit cell added would introduce more losses into the periodic structure, thus after a certain number of $N$, losses would suppress the DBE resonance. Fig. 2(b) shows the comparison between measured and simulated results. The 6-unit cell structure provides a pronounced DBE resonance in the proximity of the actual DBE frequency. The simulation results (solid lines) were obtained from the Keysight ADS Method of Moments Momentum Microwave simulator. Measured results are shown as dashed lines. The parameter magnitude $|S_{31}|$ is shown in blue while $|S_{11}|$ is shown in red, representing the transmission and reflection coefficients (of Port 1) respectively. The agreement between the measured and simulated results is excellent right around the DBE resonance frequency, $f_{d,r}$ of 2.86 GHz. A similar trend where the simulated and measured results had the best agreement around the DBE frequency was also shown in Fig. 1(c). There is a dip in $|S_{11}|$ and a peak in $|S_{31}|$, indicating the DBE resonance. Above 2.86 GHz, up to around 3.3 GHz, $|S_{31}|$ is very small confirming the formation of bandgap first noted in Fig. 1(c).

It is worth noting that $f_{d,r} < f_d$, because the DBE condition is observed in an infinitely-long periodic structure, and thus $f_{d,r}$ is in the passband and will get closer to $f_d$ as $N$ grows [16]. Even for just 6 cells, $f_{d,r}$ is only 70 MHz away from $f_d$. Since our oscillator consists of a finite number of unit cells, $|S_{31}|$ and $|S_{11}|$ show the resonance $f_{d,r}$ near the DBE frequency $f_d$ and right before the bandgap. Indeed, $f_{d,r}$ is the most important resonance because this is the frequency where the beneficial properties of DBE are seen, such as enhancement of the quality factor, as discussed in detail in [15], [16], and other oscillating features discussed next. With the information obtained from studying a passive periodic structure, we anticipate that with a proper active device configuration, the structure will oscillate at the frequency of 2.86 GHz, i.e. the resonance frequency closest to the degenerate bandgap.

### III. Performance of the DBE Oscillator

We use the same analysis found in [19], [20] to determine the ideal location for the active device, which is required to compensate for the losses and meet the Barkhausen stability criteria. By looking at the input admittance at various nodes of the periodic structure, it is found that the center of the upper line, between the 3rd and 4th cell, has the lowest real input admittance (with zero imaginary part), which is 1.2 mS for the 6-cell structure shown in Fig. 2(a) at $f_{d,r}$. The real part of the input admittance is directly proportional to the amount of negative conductance, $-g_m$, needed from the active device to cause oscillations.

We first analyze an ideal simulation setup where the active device of Fig. 1(a) is a simple voltage controlled current source with cubic behavior as

$$i(t) = -g_m v(t) + \alpha v^3(t), \qquad (1)$$

where $v(t)$ and $i(t)$ are the voltage and current of the source, $\alpha$ is a voltage amplitude saturation constant, and $-g_m$ is the negative conductance produced by the active device. An example of an I-V characteristic curve based on Eq. (1) is shown in Fig. 3(a), where the negative conductance is seen when the slope is negative.

The periodic structure consisting of 6 unit cells is simulated in the Keysight ADS Method of Moments Microwave simulator with a very fine mesh of wavelength/80. Such a fine mesh is required to ensure that the very sensitive DBE resonant behavior is captured properly. We first look at the simulated input admittance, $Y_{in}$, of the 6-cell structure, at the middle of the upper line. Naturally, the structure will oscillate at the frequency where the imaginary part of $Y_{in}$ crosses a zero, and the real part is the smallest. Fig. 3(b) shows a zoomed-in plot of real and imaginary parts of $Y_{in}$ around $f_{d,r}$. We can see that at $f_{d,r}$ there is a zero crossing of the imaginary part with a real part of 1.2 mS. Using Keysight ADS circuit simulator, the ideal active device is connected to the middle of the 6-cell structure, between cells 3 and 4 of the upper line of Fig. 2(a), and is configured to produce −1.5 and −2 mS of negative conductance $-g_m$. The voltage saturation constant $\alpha$ is chosen to be $g_m/3$; i.e., it is equal to 0.5 and 0.67 mS for the two values of $g_m$, respectively. The 6-cell structure is composed of 6 S-parameter blocks, with each block related to an s4p file obtained from ADS Microwave simulation of a single unit cell of Fig. 1(a), a dispersion diagram of which is plotted in Fig. 1(c). To confirm the presence of oscillation, transient analysis on the setup explained above is done in the Keysight ADS time-domain circuit simulator, and voltage at a 50 Ω load connected to P3 of Fig. 2(a) is carried out. Once the voltage at the 50 Ω load reaches steady-state oscillation, a time-to-frequency transformation is done on the oscillating signal using Keysight ADS's built in function to produce the spectrum shown in Fig. 3(c) for two values of the chosen $g_m$. As expected, a single, fundamental oscillation frequency of 2.86 GHz is observed for each $g_m$, confirming that the 6-cell frequency selective structure chooses $f_{d,r}$ as the oscillation frequency. This predictability of the oscillatory behavior is one of the great advantages of a structure with DBE resonance. An important aspect must be noted: the DBE occurs in a lossless waveguide [20], therefore to retain the degeneracy properties associated to it and to the DBE resonance the circuit should not be perturbed with large gain, and that is the reason we have selected a small value of −2 mS for $-g_m$.

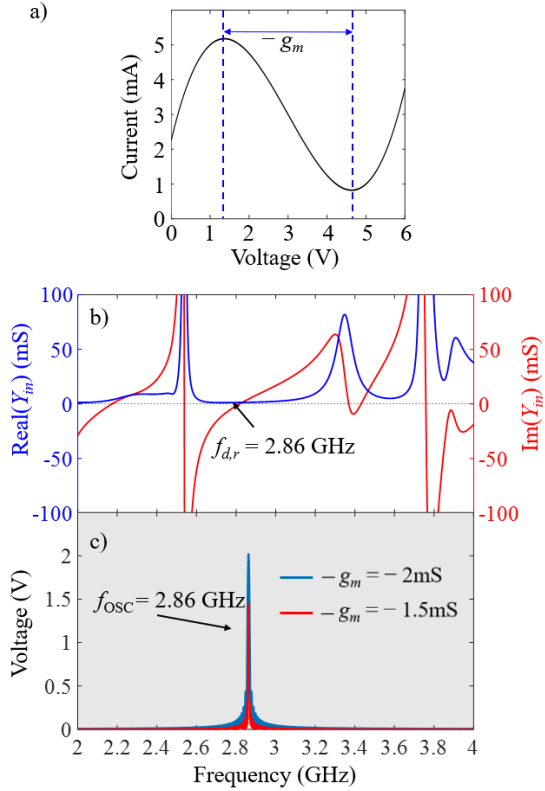

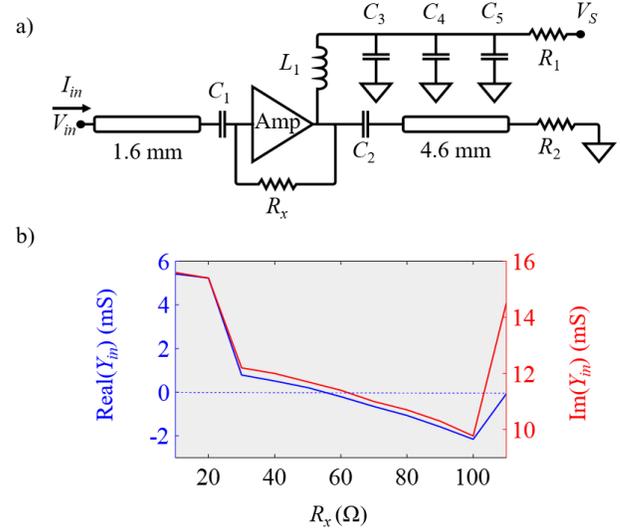

As mentioned previously, by finding the real part of the input admittance at the center of the upper line, an estimation of the amount of negative -$g_m$ needed is determined, which can be seen from Fig. 4(b), the chosen active device configuration can provide.

Fig. 3. (a) I-V characteristic of an ideal active device based on Eq. 1 producing negative conductance, $g_m$, in the region with negative slope. (b) Simulated real and imaginary part of the input admittance, $Y_{in}$, at the center of the upper line of a 6-cell structure showing a zero crossing of the imaginary part with a very small real part of 1.2 mS. (c) Simulated frequency response at the 50 Ω load of the 6-cell structure with a cubic ideal active device producing 1.5 mS and 2 mS of negative conductance, showing stable oscillation frequency of 2.86 GHz for both cases.

In the realized design, the active device is implemented using an op-amp with a feedback resistor. The schematic representation of the circuit, based on NXP Semiconductors BGA6589 MMIC power amplifier, is shown in Fig. 4(a). This simple configuration can produce a negative conductance at the op-amp input terminal, the value of which is controlled by the feedback resistor $R_x$. To measure the amount of negative conductance produced by the active device at $f_{d,r}$, a separate, one port circuit was built and its $|S_{11}|$ was measured using a Keysight E5071C Network Analyzer and later converted to admittance in Keysight ADS, which is plotted in Fig. 3(b). The active device was configured and biased as suggested by the manufacturer at the desired frequency where $C_1=C_2=47$ pF, $C_3=17$ pF, $C_4=1$ nF, $C_5=1$ μF, $L_1=1$ nH, $R_1=15$ Ω at $V_S$ of 6V, and $R_2=50$ Ω [35]. The lengths of the two transmission lines at the input and the output are 1.6 mm and 4.5 mm, respectively. From $R_x$ of 55 Ω to 100 Ω, the active device produces linear and negative conductance, shown in blue, peaking at -2 mS at $R_x$ of 100 Ω, with a susceptance of 9 mS shown in red. The susceptance of 9 mS translates to 0.5 pF equivalent shunt capacitance at $f_{d,r}$ which would have a minimal impact on the performance of the oscillator.

Fig. 4. (a) The active device configuration based on NXP Semiconductors BGA6589. Negative admittance is seen at the input. (b) Measured input admittance, defined as $I_{in}/V_{in}$, of the active device versus feedback resistor $R_x$ at $f_{d,r}$ of 2.86 GHz, showing the region with negative admittance.

We have connected the active device as shown in Fig. 5(a), and using 100 Ω as $R_x$ to provide $-g_m$ of $-2$ mS. The signal is measured at Port 3 terminated by 50 Ω, while all other ports are shorted to the ground. The frequency response of the oscillator measured at the SMA connector of Port 3 is shown in Fig. 5(b). The frequency spectrum was measured on a Keysight N9020A MXA signal analyzer with the input impedance of 50 Ω connected to Port 3 (note that 50 Ω is also the characteristic impedance of the upper microstrip, when not coupled to the bottom one). The measured spectrum's frequency matches the DBE resonance frequency, $f_{d,r}$, found in Fig. 2(b), confirming that the structure is oscillating at the predicted DBE resonance condition. Due to the strong DBE resonance, most of the energy is concentrated towards the middle of the structure [16], [36], where the active device is connected, and very little of that energy is seen at the edges. This unique energy distribution is the reason why any perturbation at the edges of the structure, where a load would likely be connected, would not greatly affect the oscillator behavior. This is one of the main advantages of this oscillator scheme over a conventional LC tank oscillator [6], [21]; the need to have a buffer. But because of that, the power delivered to the load is not very high, which may lower the efficiency of the oscillator. In this first demonstration of a DBE oscillator, the active device circuit consumes 19 dBm of DC power at the suggested biasing condition [35], while only $-7.44$ dBm of RF power is delivered to the spectrum analyzer's 50-Ω input impedance. Note that the unit cell of Fig. 1(a) was designed with a top microstrip having a characteristic impedance of 50 Ω, but the cavity is actually mismatched to the 50 Ω load because the Bloch impedance is indeed described by a 2x2 impedance matrix as described in Appendix C of Ref. [22].

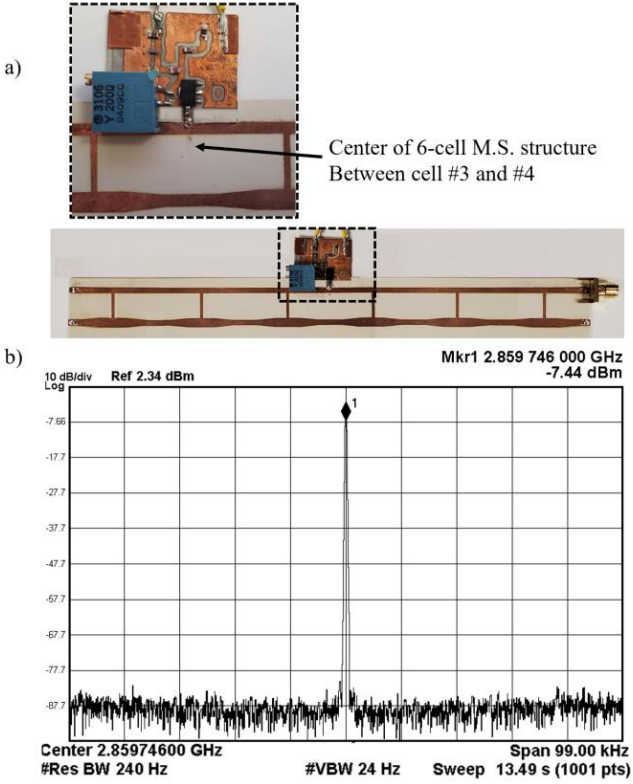

Fig. 5. (a) 6-cell cavity with the BGA6589 circuit attached to the middle of the upper line, inset showing the zoom-in of the amplifier with feedback resistor to generate negative admittance. All of the ports, except P3 are grounded; (b) Frequency domain response as seen on the spectrum analyzer with 50 Ω input impedance when connected to P3 with all other ports shorted to the ground. A fundamental frequency of oscillation of 2.859 GHz, i.e., at the expected 2.86 GHz, is observed. Resolution bandwidth is set to 240 Hz while video bandwidth set to 24 Hz to fully capture the spectrum.

Next, we investigate how robust the produced oscillation frequency is to perturbations. Perturbations may arise for several reasons, such as variation in $g_m$, the output load impedance, and even the placement of the active device. A common source of perturbation is the change of the output resistance. To emulate such a perturbation, a potentiometer was added in series with the output SMA connector on Port 3, on the upper line of the 6th cell, to add series resistance to the input impedance of the spectrum analyzer, as shown in Fig. 6(a). The overall length of the upper line has not changed, with just a small cutout in the upper microstrip line to place the potentiometer next to the SMA connector. The value of the potentiometer, denoted as $R_{POT}$, is swept from 0 to 350 Ω while power and frequency of oscillation are measured on a spectrum analyzer. The total resistance, $R_{TOT}$, seen by the 6-cell DBE structure is the sum of the potentiometer value and internal impedance of the spectrum analyzer, given by

$$R_{TOT} = R_{POT} + 50\,\Omega. \quad (2)$$

Then, the total output power is given by

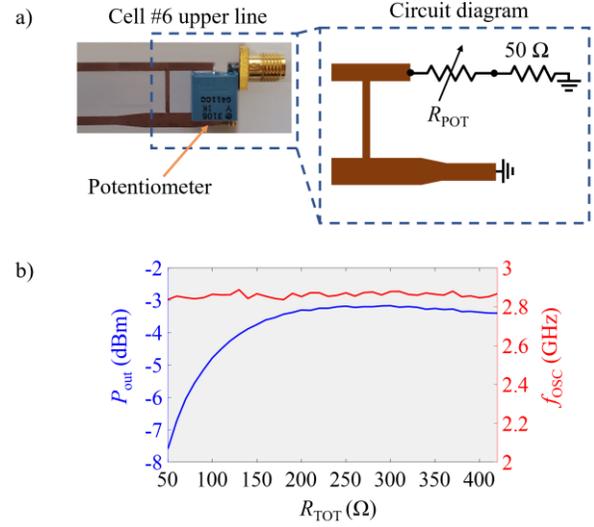

Fig. 6. (a) Connecting potentiometer in series to the SMA connector to vary the perturbing resistance $R_{TOT}$. 50 Ω is the internal impedance of the spectrum analyzer (S.A.); (b) Output power as defined by Eq. (4) (in blue) and oscillation frequency measured directly by the spectrum analyzer (in red) versus total perturbing resistance $R_{TOT}$ as defined by Eq. (2) showing the stability of oscillation frequency

$$P_{OUT} = P_{POT} + P_{S.A.}. \quad (3)$$

This is the sum of power dissipated over the potentiometer, $P_{POT}$ and the one dissipated (and measured) on the spectrum analyzer, $P_{S.A.}$. It can be rewritten as

$$P_{OUT} = P_{S.A.}\left(1 + R_{POT}/50\,\Omega\right) \quad (4)$$

where 50 Ω is the internal impedance of the spectrum analyzer. Once the measurement provides $P_{S.A.}$ for varying $R_{POT}$, then $P_{OUT}$ is calculated by (4). Also, the oscillating frequency $f_{OSC}$ is directly measured by the spectrum analyzer. The oscillation frequency and output power are both plotted in Fig. 6(b) as a function of the total load resistance. This figure shows that while the total perturbing resistance $R_{TOT}$ varies from 50 to 400 Ω, the oscillation frequency is very stable around the value of 2.86 GHz. When $R_{TOT}$ is 50 Ω, i.e., when $R_{POT} = 0$ Ω, $P_{OUT}$ is the same value as seen in Fig. 5(b). Then, the output power grows for increasing $R_{TOT}$ up to 250 Ω, and decreases slightly as $R_{TOT}$ increases further. Fig. 6(b) shows the resilience of the DBE oscillator to changes in the load values.

An important characteristic of any oscillator is its ability to produce a signal with near-perfect periodicity in the time domain, i.e. its phase noise. Fig. 7(a) shows the wideband spectrum of the oscillator measured at the spectrum analyzer with 50 Ω internal impedance (the series potentiometer used to perturb the $R_{TOT}$ is not present for this measurement, i.e. $R_{POT}$ is 0). Phase noise is shown in Fig. 7(b) for measured and simulated results. For the simulated results, two values of $-g_m$ were used, −1.8 mS and −2 mS. To get the best agreement between measured and simulated results, $g_m$ was reduced to −1.8 mS. Since an idealized simulation bench in Keysight ADS was used, a small level of disagreement is expected

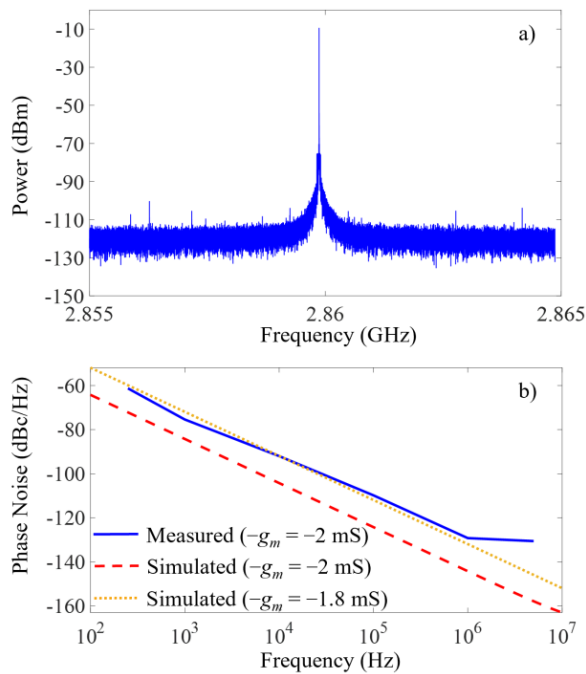

Fig. 7. (a) Measured wideband spectrum at the 50 Ω load terminal of the Keysight N9020A MXA signal analyzer and plotted using Mathworks Matlab; (b) Plot of measured and simulated phase noise at frequency offsets from 100 Hz to 100 MHz. Simulated results are plotted for two values of $-g_m$: −1.8 mS and −2 mS.

discrepancy is expected.

The measured phase noise was calculated based on the measured power spectrum results with the help of Mathworks Matlab Mixed Signal Blockset library and phaseNoiseMeasure function specifically [37]. The Matlab function automatically calculates the single noise sideband (SSB) at specified offsets from the ideal oscillation frequency. Matlab library was used since Keysight N9020A MXA signal analyzer used to measure the spectrum does not have the module installed capable of directly measuring the phase noise. The simulated phase noise was obtained from the Keysight ADS Harmonic Balance simulator by considering an added rms noise voltage source to the ideal active device described in Section III. The variance of the noise voltage source is proportional to $kT/g_m$, where $k$ is the Boltzmann constant, T is the temperature (assumed to be 293 K), and $-g_m = -1.8$ mS and −2 mS is the value of negative conductance provide by the ideal active device. The agreement between the simulated and measured results; besides that, the experimental negative conductance may slightly differ from −2 mS when the active circuit is mounted on the waveguide. As mentioned previously, to obtain the value of negative conductance produced by the active device at a certain value of feedback resistance, a separate 1-port circuit was built and measured using a VNA. In that configuration, the active device sees the input impedance of the VNA, which is 50 Ω, while when connected to a 6-cell periodic structure, it does not see exactly 50 Ω. Another reason why the active device circuit potentially does not produce precisely −2 mS at the right biasing is fabrication imperfections. Any imperfections in the microstrip will lead to changes in impedance and increase the losses. And yet, the general agreement depicted in the plot is a further testament to the predictability of the behavior of the proposed oscillator. These results are only slightly higher as compared to a more conventional LC oscillator, while having a much more complex geometry/implementation [38], [39].

## IV. CONCLUSION

A first-ever fabricated oscillator based on a periodic coupled-microstrip waveguide with a DBE condition is presented. It shows a very stable oscillation behavior near the DBE frequency, even for a wide range of output impedance variation, which is consistent with the resonant behavior of the passive degenerate waveguide. The oscillation frequency is associated with the DBE resonance frequency which can be controlled by the DBE structure's dimensions and substrate characteristics. The study presented, based on both simulated and measured results, shows that this oscillator design is promising as an RF source, where a stable oscillation frequency is of high importance. Future work should focus on the robustness of the oscillator structure to structure's perturbations as well as employing other active device structures/configurations.

## V. ACKNOWLEDGMENT

This material is based upon work supported by the National Science Foundation under award NSF ECCS-1711975.